\documentclass{icrc2009}

\usepackage{graphicx}   
\usepackage[caption=false]{caption}    
\usepackage[font=footnotesize]{subfig} 
\usepackage{fixltx2e}
\usepackage{url}

\newcommand{\shorttitle}[1]%
{\markboth{Proceedings of the 31\MakeLowercase{$^{st}$} ICRC, {\L}\'{o}d\'{z} 2009}{#1} }

\begin{document}
\title{Unidentified Gamma-Ray Sources as Ancient Pulsar Wind Nebulae}

\author{\IEEEauthorblockN{O.C. de Jager\IEEEauthorrefmark{1},
S.E.S. Ferreira\IEEEauthorrefmark{1}, 
A. Djannati-Ata\"{i}\IEEEauthorrefmark{2},
M. Dalton\IEEEauthorrefmark{3},
C. Deil\IEEEauthorrefmark{4},
K. Kosack\IEEEauthorrefmark{5},
M. Renaud\IEEEauthorrefmark{2}, \\
U. Schwanke\IEEEauthorrefmark{3} and 
O. Tibolla\IEEEauthorrefmark{4}}
                            \\
\IEEEauthorblockA{\IEEEauthorrefmark{1}Unit for Space Physics, North-West University, Potchefstroom, South Africa}
\IEEEauthorblockA{\IEEEauthorrefmark{2}Astroparticule et Cosmologie, Universite Paris 7 Denis Diderot, Paris, France}
\IEEEauthorblockA{\IEEEauthorrefmark{3}Institut f¨ur Physik, Humboldt-Universit¨at zu Berlin, Berlin, Germany}
\IEEEauthorblockA{\IEEEauthorrefmark{4}Max-Planck-Institut f¨ur Kernphysik, Heidelberg, Germany}
\IEEEauthorblockA{\IEEEauthorrefmark{5}CEA Saclay, F-91191 Gif-sur-Yvette Cedex, France}
}

\shorttitle{de Jager et al.: Ancient Pulsar Wind Nebulae}
\maketitle

\begin{abstract}
In this paper we explore the evolution of a PWN while the pulsar is spinning down.
An MHD approach is used to simulate the evolution of a composite remnant. Particular
attention is given to the adiabatic loss rate and evolution of the nebular field strength
with time. By normalising a two component particle injection spectrum (which can reproduce the
radio and X-ray components) at the pulsar wind termination shock to the 
time dependent spindown power, and keeping track with losses since pulsar/PWN/SNR birth,
we show that the average field strength decreases with time as $t^{-1.3}$, so that
the synchrotron flux decreases, whereas the IC gamma-ray flux increases, until most of
the spindown power has been dumped into the PWN. Eventually adiabatic and IC losses will
also terminate the TeV visibility and then eventually the GeV visibility.
\end{abstract}
 
\section{Introduction}
Aharonian et al. \cite{ahar:08} discussed eight unidentified VHE gamma-ray sources 
discovered with the H.E.S.S. telescopes. All are extended
objects with angular sizes ranging from approximately 3
to 18 arc minutes, lying close to the Galactic plane (suggesting
they are located within the Galaxy). In each case, the spectrum
of the sources in the TeV energy range can be characterized as
a power-law with a differential spectral index in the range 2.1
to 2.5. The general characteristics of these sources (spectra,
size, and position) are similar to previously identified galactic
VHE sources (e.g. pulsar wind nebulae PWNe), however since these sources
have so far no clear counterpart in lower-energy wavebands,
further multi-wavelength study is required to understand the
emission mechanisms powering them, and therefore follow-up
observations with higher-sensitivity X-ray and GeV $\gamma$-ray
telescopes will be beneficial (as stated in \cite{ahar:08}.)

One possibility is that we are dealing with relatively old PWN
born from Type II supernovae, but still relatively close to the
molecular clouds from which the massive progenitor stars were born. 
This will then also explain their proximity to the galactic plane.

A natural explanation would be that these sources were once bright
in synchrotron emission, but that the field strength decreased with time
as the PWN expanded with time \cite{dejager}: The pulsar eventually deposited all its spindown power
into the nebula and whereas the synchrotron brightness decreased with time
because of field decay, the inverse Compton $\gamma$-ray flux increases until 
reaching a convergent value, after which it will also decay because of
continuous adiabatic losses and inverse Compton cooling. The $\gamma$-ray
lifetime of a PWN can then be much longer than the apparent radio and X-ray lifetimes.

In this paper we will discuss the results of MHD and radiative modelling
of evolving PWNe and show predicted evolutionary results for the composite
SNR G21.5-0.9.
 
\section{The MHD model for composite supernova remnants}
Supernova remnant evolution in either uniform or non-uniform media have been modelled extensively by e.g. \cite {Tenorio-Tagle-etal-1991,Jun-and-Jones-1999}. For either composite SNRs or PWNe in the ISM simulations were also presented by e.g.\cite{Van-der-Swaluw-etal-2001,Bucciantini-2002,Bucciantini-etal-2003, van-der-Swaluw-2003,Del-Zanna-etal-2004,Bucciantini-etal-2005}. In this work we use a similar model as used in most of the studies above by solving the well known Euler equations
\begin{eqnarray}\label{hd}
\frac{\partial}{\partial t}\rho+\nabla\cdot(\rho\textbf{v})=0, \\
\frac{\partial}{\partial t}(\rho\textbf{v})+\nabla\cdot(\rho\textbf{v}\textbf{v}+P\textbf{I})=0, \\
\frac{\partial}{\partial t}(\frac{\rho}{2}\textbf{v}^2+\frac{P}{\gamma-1})+\nabla\cdot(\frac{\rho}{2}\textbf{v}^2\textbf{v}+
\frac{\gamma\textbf{v}P}{\gamma-1})=0
\end{eqnarray}
which describe inviscid flow. Here $\rho$ is the density, $\textbf{v}$ the velocity and $P$ the gas pressure.  These equations describe the balance of mass, momentum and energy. Currently we only consider a one fluid scenario with an adiabatic index of 5/3. Although a relativistic description is necessary to model PWN evolution correctly, the speed of the relativistic material downstream of the pulsar wind termination shock is sufficiently smaller than $c$ to use a non-relativistic treatment (see also e.g. \cite{Van-der-Swaluw-etal-2001}). The numerical scheme is discussed in \cite{Leveque-2002} and compute solutions to hyperbolic differential equations using a wave propagation approach. See also \cite{Ferreira-and-de-Jager-2008} for more details. The model solves in spherical coordinates $r$ and $\phi$, with $r$ ranging from 0.01 pc to 25 pc (2000 gridpoints) and $\phi$ from $0^{o}$ to $180^{o}$ (150 gridpoints).

For the initial and boundary conditions of the SNR (see also \cite{Blondin-and-Ellison-2001,Van-der-Swaluw-etal-2001,Bucciantini-etal-2003, Del-Zanna-etal-2004})
we assume a spherical region, radius $r_{ej}$, 
and a high constant density $\rho_{ej}$ with a radially increasing velocity profile
\begin{eqnarray}
v=\frac{r}{t}=v_{ej}r/r_{ej}. 
\end{eqnarray}
In this case we take $r_{ej}=0.1$ pc while for the density we have
\begin{eqnarray}\label{density}
\rho_{ej}=\frac{3M_{ej}}{4 \pi r_{ej}^3} 
\end{eqnarray}
with $M_{ej}$ the ejecta mass. For the velocity we have 
\begin{eqnarray}\label{speed}
v_{ej}=\sqrt{\frac{10}{3}\frac{E_{ej}}{M_{ej}}}.
\end{eqnarray}

To compute the PWN magnetic field we solve 
\begin{eqnarray}
\frac {\partial \textbf{B}}{\partial t} + \nabla \times (\textbf{v} \times \textbf{B}) =0
\end{eqnarray}
using a similar scheme as for the fluid part. Note that this is not a full MHD solution because the field is calculated kinematically from the flow (\cite{Scherer-and-Ferreira-2005a, Ferreira-and-Scherer-2006}) and no backreaction on the fluid is considered. More detailed MHD calculations were done by e.g \cite{van-der-Swaluw-2003} \& \cite{Bucciantini-etal-2005}.

\section{Constraints from pulsar evolution}

For the PWN we assume that the spin-down luminosity of the pulsar is given by
(assuming a pulsar braking index of 3)
\begin{eqnarray}\label{lum}
L(t)=\frac{L_{0}}{\left( 1+\frac{t}{\tau}\right)^2},
\end{eqnarray}
where $L_0$ is the initial spindown power and $\tau$ the spindown timescale, which, 
for a birth period $P_0$ and present period $P$, is defined as
\begin{eqnarray}
\tau=\frac{2\pi^2 I}{P_0^2L_0}=\frac{2\pi^2 I P_0^2}{P^4L}.
\end{eqnarray}

\section{The evolution of the plerionic magnetic field strength}

To calculate the multiwavelength (MWL) spectrum we need to know the behaviour of the average PWN field strength $B(t)$ with time. This quantity was calculated by taking the volume averaged field strength between
the pulsar wind termination shock radius and PWN outer radius.

The calculation of the average field strength starts progressively later (in time) with decreasing
$L_0$. This is because of the difficulty in resolving the position of the PWN termination
shock as $L_0$ decreases. This difficulty should be resolvable if we reduce the grid
size of the calculation, but at the expense of CPU calculation time. For example, the
PWN termination shock radius of G21.5-0.9 is $\sim 0.5$ arcsec, corresponding
to a shock radius of 0.01 pc, which is already consistent with the minimum assumed
grid size. 

\begin{figure}
  \includegraphics[width=3.0in]{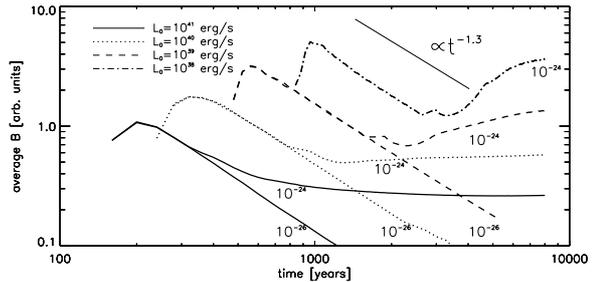}
  \caption{The average magnetic field strength of the PWN for $\tau=3000$ y and
$L_0$ in the range as indicated. Two values of the ISM density was assumed. The slope 
$\propto t^{-1.3}$ indicates the approximate pre-reverse shock field decay evolution
with time. \label{fig:bavga}}
\end{figure}

Figure \ref{fig:bavga} shows the behaviour of the average $B(t)$ 
for $L_0$ ranging between $10^{38}$ to $10^{41}$ erg/s and ISM densities of $10^{-26}$ g/cm$^{3}$
and $10^{-24}$ g/cm$^{3}$. A more detailed discussion of this will be given elsewhere.

Prior to the passage of of the reverse shock, we find that the field strength 
decreases as $t^{-1.3}$, independent of the chosen parameters. This is modified by
the reverse shock, but after passage, 
the time evolution is expected to revert back to this $t^{-1.3}$ behaviour.

\section{Adiabatic losses}
In this section the evolution of a PWN inside a SNR is studied. Model solutions corresponding to $M_{ej}=8M_{\odot}$ in Equation \ref{density} and spin-down time $\tau$ = 3000 y and $\tau$ = 300 y in Equation \ref{lum} are shown . Different scenarios ranging from initial pulsar wind luminosity $L_{0} = 10^{41}$ erg/s to $L_{0}=10^{38}$ erg/s in Equation \ref{lum} are shown. 

The rate of change of the energy of a particle convected by a pulsar wind expanding at a velocity $\vec{V}$
is given by
\begin{eqnarray}\label{dedt}
\frac{dE}{dt} = -\frac{E}{3}(\nabla\cdot \vec{V})
\end{eqnarray}
Below we will see that this quantity is expected to be negative, giving rise to adiabatic losses, unless the PWN is sufficiently crushed by the reverse shock, such that
the term $\nabla\cdot \vec{V}<0$, in which case the particles will experience adiabatic heating.
For practical purposes we calculate the average adiabatic energy loss rate
over the PWN between the termination shock and PWN radii by averaging the quantity
$\nabla\cdot \vec{V}$ over volume. The radius of the PWN was determined by establishing the
position where the PWN field strength drops to zero.
 \begin{figure*}[!t]
   \centerline{\subfloat[$\tau=300$ yr]{\includegraphics[width=2.5in]{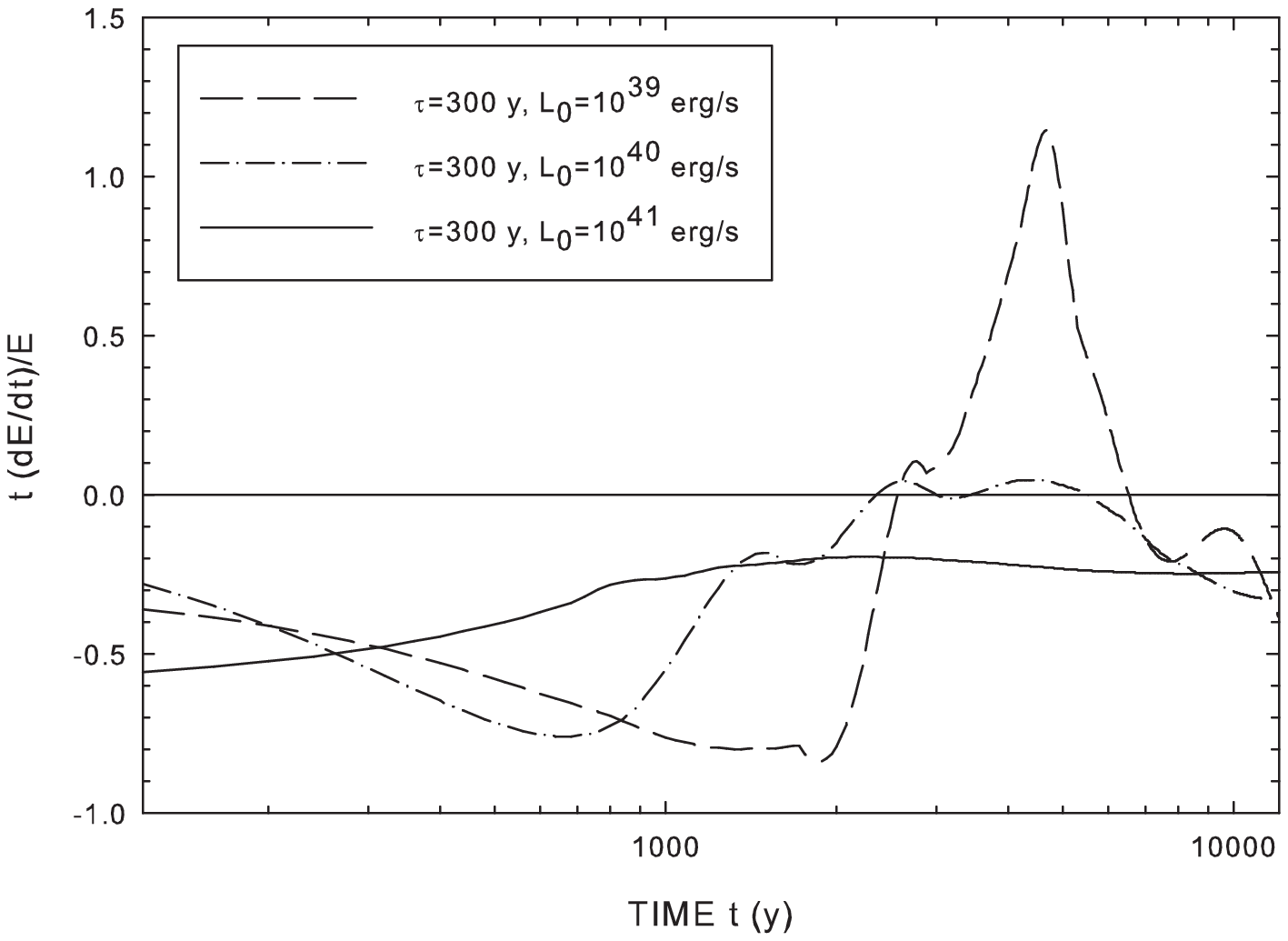} \label{dedt300}}
              \hfil
              \subfloat[$\tau=3000$ yr]{\includegraphics[width=2.5in]{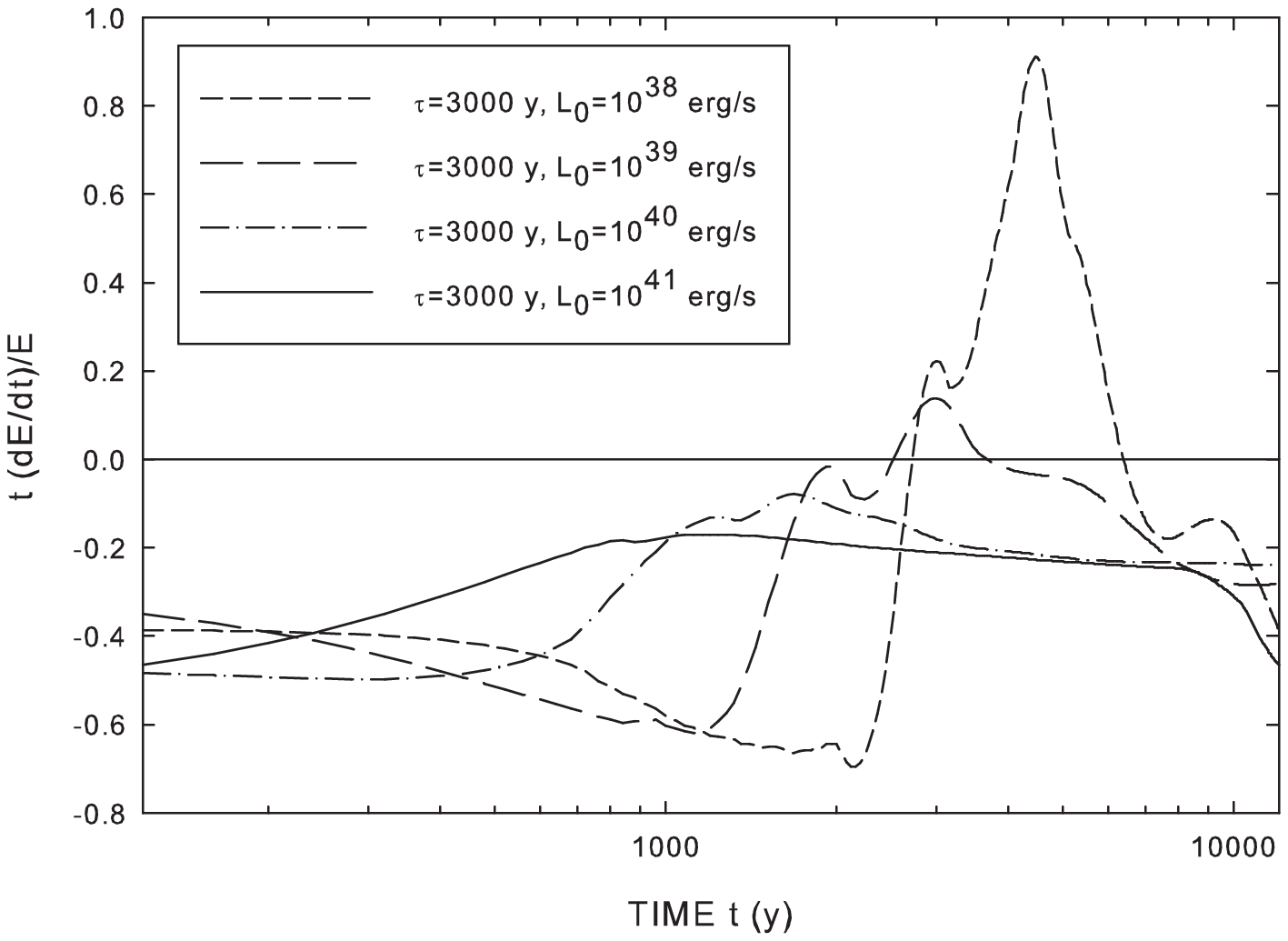} \label{dedt3000}}
             }
   \caption{The scaled relative energy loss rate  $t(dE/dt)/E$ (a dimensionless quantity)
due to adiabatic expansion as a function of time $t$ since birth. The spindown timescale in this case is 
$\tau=300$ y (left) and 3000 y (right), whereas the SNR ejecta mass for both cases
was $8M_{\odot}$. The spindown power $L_0$ at birth is indicated in the Legend.}
   \label{adiabatic}
 \end{figure*}
We scale the abovementioned rate of energy change (averaged over volume) by multiplying the relative energy loss rate
$(dE/dt)/E$ with the age $t$ of the PWN to give the dimensionless quantity $t(dE/dt)/E$. The results
are shown in Figures \ref{dedt300} and \ref{dedt3000} for spindown timescales of $\tau=300$ y and
$\tau=3000$ y respectively and PSR/SNR parameters discussed above.

Initially we find that the quantity $t(dE/dt)/E$ is negative as a result of expansion, so that the particles loose
energy due to this process. However, when the reverse shock compresses the PWN, we find that the quantity
$\nabla\cdot \vec{V}$ becomes negative, in which case the particles will start to gain some of their lost energy.
With decreasing $L_0$ we find that this heating process starts at progressively later times, since the
time when the reverse shock encounters the PWN increases with such decreasing $L_0$ for the same $\tau$.

It is interesting to note that the quantity  $|t(dE/dt)/E|$ is always less than unity
except when the reverse shock compresses the flow. 

Since the relative adiabatic energy loss rate is nearly constant at a value around -0.5 (excluding the time of
reverse shock passage), the total adiabatic losses of a particle injected during birth 
with initial energy $E_0$ and which can survive significant radiation losses would be
\begin{eqnarray}\label{ad}
E \sim E_0\left(\frac{t}{t_0}\right)^{-0.5}.
\end{eqnarray}
Note that $E=0$ if $t_0=0$, which implies an inconsistent solution, unless $t_0>0$.
We find that $t_0\le 100$ yr but we are currently reducing the grid size of the simulation and will report on the solution
for a convergent value of $t_0$ in a followup paper.

\section{Time dependent evolution of the lepton spectrum}

We define $N(E,t)$ as the time dependent differential particle spectrum for leptons of energy $E=\gamma m_ec^2$, whereas
$\tau_{\rm syn}$ and $\tau_{\rm ad}$ are the timescales corresponding to synchrotron and adiabatic
losses respectively. The magnetic field strength $B(t)$ (used in $\tau_{\rm syn}$) is time dependent. We then integrate the transport equation 
\begin{equation}
\frac{dN}{dt}+\frac{N}{\tau_{\rm syn}}+\frac{N}{\tau_{\rm ad}}=Q(t)
\end{equation}
between time $t=0$ when $P=P_0$, i.e. the pulsar birth period and the 
current epoch at $T_{\rm SNR}$ assuming a pulsar braking index $n=3$. 
From $N$ we calculate the spectral energy distributions (SED) in synchrotron and inverse Compton as discussed below.

We adopt the injection spectrum of \cite{vd06} for electrons at the pulsar wind shock
\begin{equation}\label{q}
Q(E,t)=\left(
\begin{array}{l}
Q_0(t)(E/E_b)^{-p_1}\,\,{\rm for}\,\,E<E_b\\
Q_0(t)(E/E_b)^{-p_2}\,\,{\rm for}\,\,E_b<E<E_{\rm max}
\end{array}\right),
\end{equation}
with $E_b$  the intrinsic break energy between the radio and X-ray components.
A value of $p_1 \sim 1.0$ reproduces the typical flat radio spectra, whereas $p_2\sim 2$
would reproduce the uncooled spectral indices seen in X-rays at the pulsar wind termination 
shock.

Following \cite{vd06}, the energy
equation for $Q(t)$ can be written in terms of the time dependent spindown power $L(t)$ giving
\begin{equation}\label{eta}
\int Q(\gamma,t)E dE=\eta L(t).
\end{equation}
We will assume the conversion efficiency $\eta$ of spindown power to particles as a free parameter. 
The total injected lepton energy over time $t$ since birth is then (assuming a constant $\eta$)
\begin{equation}
W_e(t)=\int_0^t \eta L(t)dt=\eta \Delta E_{\rm rot},
\end{equation}
where $\Delta E_{\rm rot}=I(\Omega_0^2-\Omega^2)/2$ (with $\Omega=2\pi/P$) is the net kinetic rotational energy deposited between
birth and time $t$.

 \begin{figure*}[!t]
   \centerline{\subfloat[X-rays]{\includegraphics[width=2.5in]{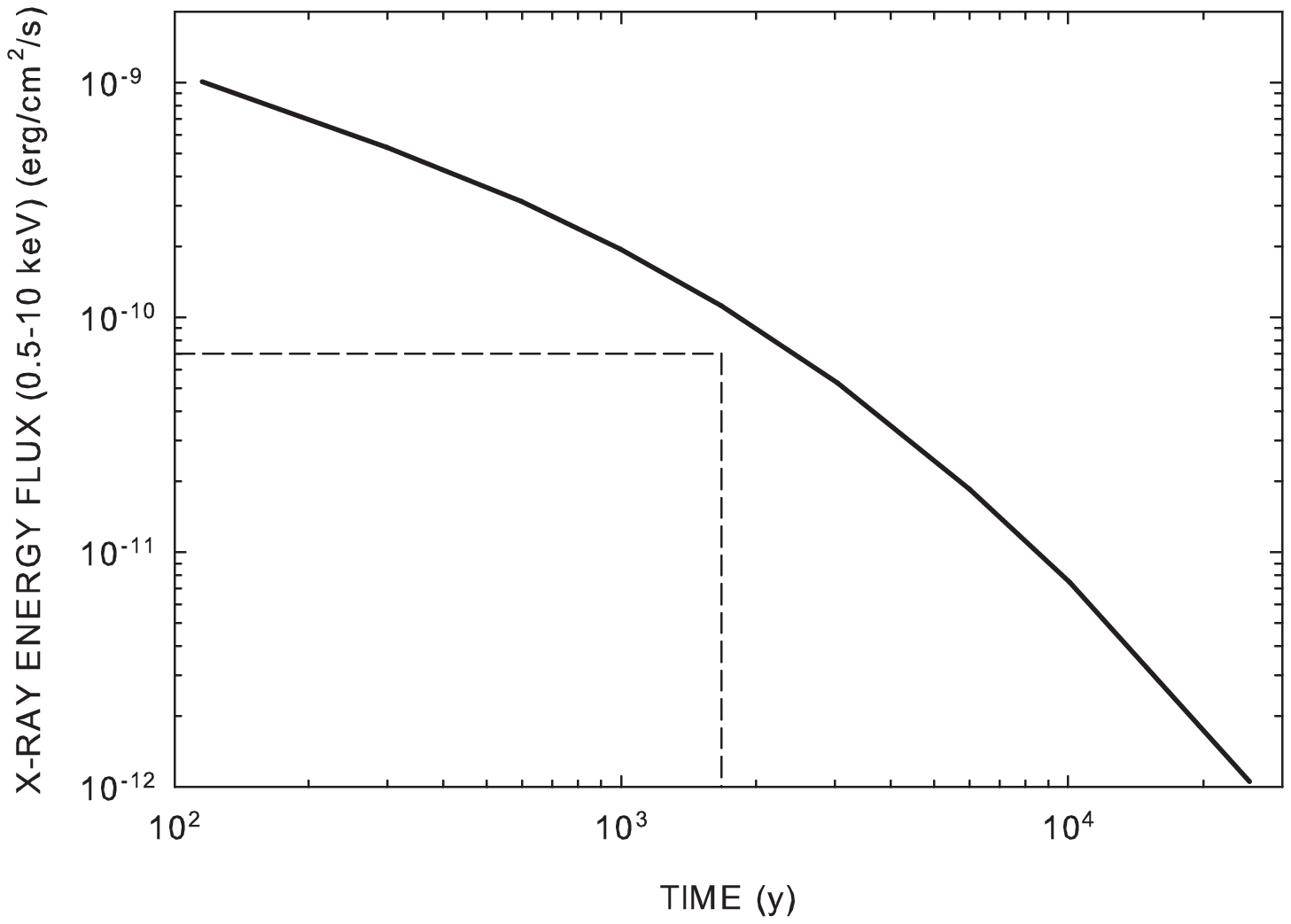} \label{x-ray}}
              \hfil
              \subfloat[GeV and TeV]{\includegraphics[width=2.5in]{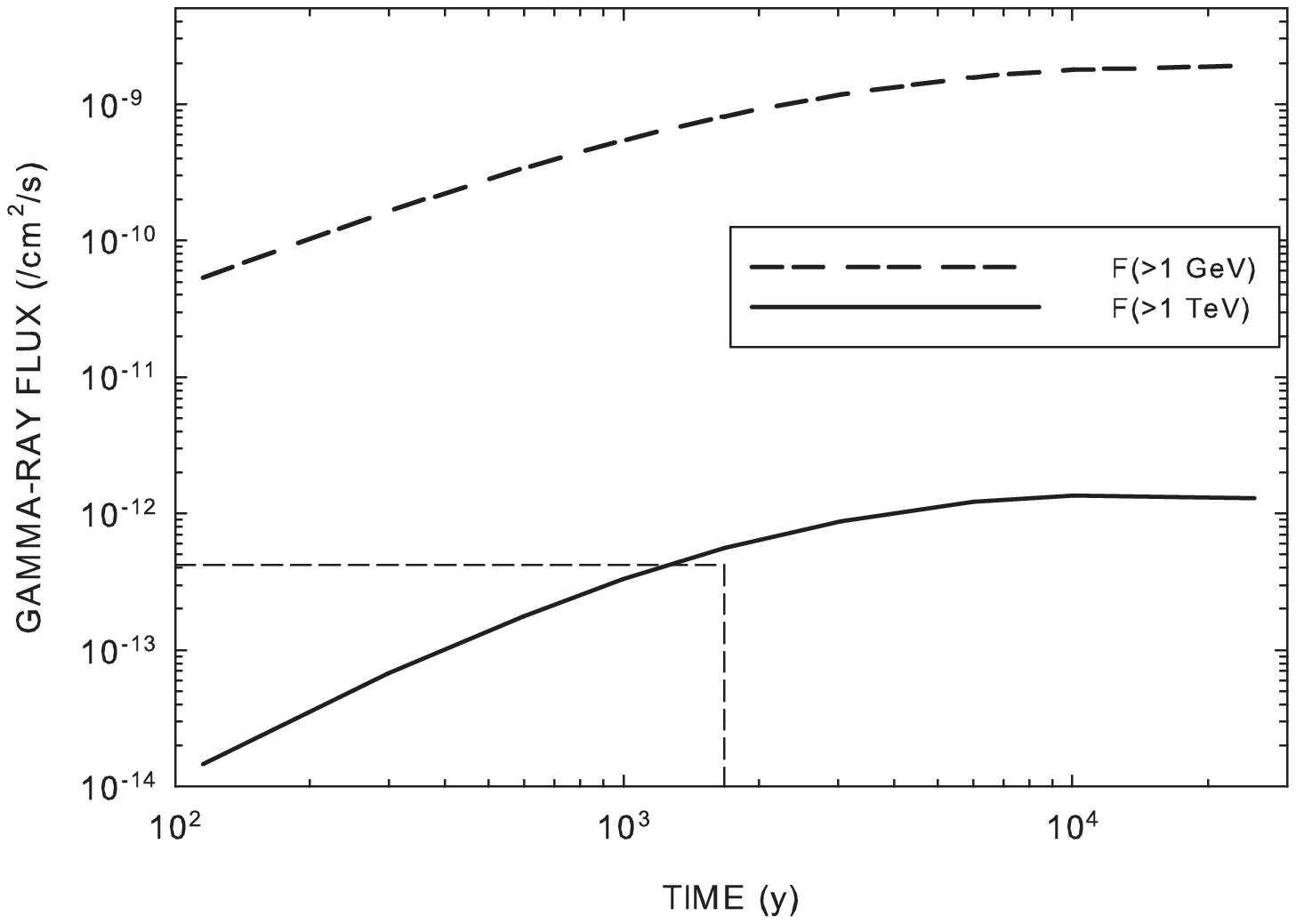} \label{gev-tev}}
             }
   \caption{The evolution of the X-ray (left) and GeV/TeV (right) fluxes with time. The
dashed lines indicate the present state of G21.5-0.9.}
   \label{flux}
 \end{figure*}

\section{Evolution towards an unidentified gamma-ray source}

In our evolutionary model we will use the young composite SNR
G21.5-0.9 as an example and follow the time evolution of the leptonic spectrum 
and hence MWL intensity. The central pulsar PSR J1833-1034 has a period of 61.8 ms
and for an expansion age near 1 kyr \cite{bieten}, the spindown timescale $\tau$ 
should vary between 3000 and 3800 y given an inferred birth period $P_0$ between
$\sim 50$ and 55 ms. The corresponding
initial spindown power ranges between $L_0=5\times 10^{37}$ and $10^{38}$ erg/s.

We will use $p_1=1$ as observed in radio \cite{salter} while
for X-rays we would expect that a value of $p_2=2$ 
corresponding to the pulsar wind termination shock \cite{slane}
would reproduce the MWL spectrum best. However, an average value of
$p_2=2.6$ seems to fit the data better.

To reproduce the ratio of energy fluxes
between X-rays and TeV, we normalise the average field strength to $25\mu$G at the present
age near 1 kyr. 

ISO observations \cite{gallant98} 
of the PWN show that the radio spectrum should break around $10^{12}$ Hz. This break is
either intrinsic or due to radiation losses. We find however that this break cannot
be due to radiation losses since this would imply a too large Crab-like field
strength, which cannot be reconciled with the observed ratio of TeV to X-ray flux.
An intrinsic break at energy $E_b$ to $p_2=2.6$ best reproduces the post spectal break data.

For a birth period of $P_0=50$ ms we still need a relatively large conversion efficiency
of $\eta=0.7$ in eqn \ref{eta}
to reproduce the observed synchrotron and IC spectra at the present epoch.
The required break energy in eqn \ref{q} is $E_b=40$ GeV, which we
will keep fixed with time since we have no theory on the evolution of $E_b$. The assumed radiation fields for the IC 
calculation were the CMBR, a 25K galactic dust component and starlight component corresponding
to 1 eV/cm$^{3}$. The latter two radiation energy densities agree approximately
with the values found for the inner galaxy region at the location of G21.5-0.9 by \cite{p06}.

Assuming a constant $\eta$ with time, but the spindown power decreasing as that of 
a magnetic dipole, and hence decreasing particle input with time, we were able to
calculate the time evolution of $Q(\gamma,t)$ and hence the MWL spectrum from which
the time evolution of the radio, X-ray and TeV fluxes were calculated. The latter two are
shown in Figure \ref{flux}. It is clear that the X-ray flux decreases with time given the
decreasing magnetic field strength with time, whereas both the inverse Compton GeV and TeV fluxes
increases with time, reaching a limiting value. The predicted radio, X-ray and TeV fluxes
agree with the observed fluxes at the present epoch.

\section{Conclusions}
In this paper we have given the basic ingredients which gives the time evolution
of the MWL spectrum of a PWN. The basic result is the following: Whereas the X-ray flux is large during
early epochs, the GeV and TeV fluxes start at relatively low values. As time progresses towards
the Vela and post-Vela epoch, the synchrotron flux starts to decrease significantly, whereas the IC flux uncreases,
until reaching a steady state value. Given the page limit of this paper, we could not explore the details of IC and adiabatic losses which would affect the time evolution at epocs $\gg 10$ kyr. This
will be discussed in a followup paper. 

The basic conclusion however remains, as a PWN grows older, it can remain bright in IC, whereas the GeV/TeV flux remains high. This can continue until IC and adiabatic losses, or, breakup and diffusion
into the ISM finally terminates the gamma-ray lifetime.

\end{document}